\documentclass[11pt]{article}
\usepackage{epsfig} 
\newcommand{\Nf}{N_{\!f}}

\begin{document}
\title{On the $n$ $\rightarrow$ $0$ limit of $\gamma_{gg}(a)$ in QCD} 
\author{J.F. Bennett \& J.A. Gracey, \\ Theoretical Physics Division, \\ 
Department of Mathematical Sciences, \\ University of Liverpool, \\ Peach 
Street, \\ Liverpool, \\ L69 7ZF, \\ United Kingdom.} 
\date{} 
\maketitle 
\vspace{5cm} 
\noindent 
{\bf Abstract.} We consider the $n$ $\rightarrow$ $0$ limit of the DGLAP 
splitting function $\gamma_{gg}(a)$ at all orders in the strong coupling 
constant, $a$, by analysing the leading order large $\Nf$ form of the 
associated $d$-dimensional critical exponent. We show that for unpolarized 
scattering the pole at $n$ $=$ $0$ which appears in successive orders in 
perturbation theory is absent in the resummed expression. 

\vspace{-17cm} 
\hspace{13.5cm} 
{\bf LTH 483} 

\newpage 

The DGLAP equation, \cite{1}, is widely used to evolve the parton structure 
functions of the nucleon constituents over a large range of energy scales. 
Central to the equation are the splitting functions which depend on the 
variable $x$ which represents the fraction of the momentum carried by that 
parton in the nucleon. Alternatively one can work with the anomalous dimension 
of the underlying twist-$2$ operators built out of the quark and gluon fields 
of QCD which depend on the variable $n$. They are related to the splitting 
functions via a Mellin transform restricted to the unit interval in $x$. It is 
generally accepted that the solution of the DGLAP evolution which effectively 
represents perturbative QCD, fits the data extremely well. See, for example,
\cite{2}. Moreover, it appears to transcend the region where the perturbative
approximation ought not to be valid. Therefore, one would hope that either by 
resumming perturbation theory in some fashion or developing non-perturbative 
methods, such as the BFKL formalism \cite{3}, it might be possible to begin 
to explore phenomena in the more extreme non-perturbative regions. One recent 
issue, \cite{4}, has been the problem of understanding the $n$ $\rightarrow$ 
$0$ behaviour of the DGLAP splitting functions in QCD. It is known that in 
perturbation theory each term of the expansion of the operator anomalous 
dimensions in the strong coupling constant has successively higher order poles 
at $n$~$=$~$0$, \cite{5,6,7}. However, other considerations suggest that the 
function is finite at $n$~$=$~$0$, \cite{4,8}, though there is some
disagreement with this point of view, \cite{9}. In this letter we provide some 
more insight into the $n$ $\rightarrow$ $0$ limit of the gluon-gluon splitting 
function, $\gamma_{gg}(a)$, where $a$~$=$~$\alpha_s/(4\pi)$ is the strong 
coupling constant. This is achieved by examining the large $\Nf$ result for 
$\gamma_{gg}(a)$ which has been computed in \cite{10} where $\Nf$ is the number
of quark flavours. Essentially the $1/\Nf$ expansion sums a different set of 
Feynman diagrams from those which are ordinarily computed in the loop or 
coupling constant expansion of perturbation theory. Moreover, the resummation 
is to all orders in $a$. This technique has been used in \cite{11,12,10,13} to 
determine the anomalous dimensions of all the twist-$2$ operators used in deep 
inelastic scattering for both unpolarized and polarized processes at 
$O(1/\Nf)$. The all orders results are expressed as a function of $d$, where 
$d$ is the dimension of spacetime, known as critical exponents. The 
coefficients of the corresponding renormalization group function are deduced 
from knowledge of the $d$-dimensional fixed point of the QCD $\beta$-function 
and properties of the critical renormalization group equation. The results have 
been shown to be in agreement with all known explicit two and three loop 
perturbative results, \cite{5,6,7,14,15}, which puts the validity of the large 
$\Nf$ results in relation to deep inelastic scattering on a firm footing. 
Hence, in this letter we will examine the $n$ $\rightarrow$ $0$ limit of the 
singlet gluon splitting function at $O(1/\Nf)$ and all orders in $a$.

First, we recall the basic formalism of the problem. As the flavour singlet
gluonic twist-$2$ operator mixes with the fermionic operator under 
renormalization in perturbation theory, one has to deal with a matrix of 
anomalous dimensions, $\gamma_{ij}(a)$. Since we will be considering it in the 
large $\Nf$ expansion we define the coefficients of its perturbative expansion 
formally by  
\begin{equation} 
\gamma_{ij}(a) ~=~ \left( 
\begin{array}{ll} 
\gamma_{qq}(a) & \gamma_{gq}(a) \\ 
\gamma_{qg}(a) & \gamma_{gg}(a) \\ 
\end{array} 
\right) 
\label{mixmat} 
\end{equation} 
where 
\begin{eqnarray}  
\gamma_{qq}(a) &=& a_1a + (a_{21}\Nf + a_{22})a^2 + (a_{31}\Nf^2 + a_{32}\Nf 
+ a_{33})a^3 + O(a^4) \nonumber \\ 
\gamma_{gq}(a) &=& b_1a + (b_{21}\Nf + b_{22})a^2 + (b_{31}\Nf^2 + b_{32}\Nf 
+ b_{33})a^3 + O(a^4) \nonumber \\ 
\gamma_{qg}(a) &=& c_1 \Nf a + c_2\Nf a^2 + (c_{31}\Nf^2 + c_{32}\Nf 
+ c_{33})a^3 + O(a^4) \nonumber \\ 
\gamma_{gg}(a) &=& (d_{11}\Nf + d_{12})a + (d_{21}\Nf + d_{22})a^2 
+ (d_{31}\Nf^2 + d_{32}\Nf + d_{33})a^3 + O(a^4) ~.  
\label{coeffdef} 
\end{eqnarray} 
The explicit values of the coefficients to two loops as a function of $n$ are
given in \cite{5,6,7} and exact values for the low operator moments at three 
loops are found in \cite{14,15}. In particular we note
\begin{equation} 
d_{11} ~=~ \frac{4}{3} T(R) 
\end{equation}  
where $T(R)$ is given by $\mbox{tr}(T^a T^b)$ $=$ $T(R)\delta^{ab}$ and $T^a$
are the generators of the colour group whose structure constants are $f^{abc}$.
In \cite{10} it was argued that the set of coefficients of the $C_2(G)$ 
sector of $\gamma_{gg}(a)$ at leading order in $1/\Nf$, corresponding to 
$d_{l1}^{C_2(G)}$ at the $l$th loop, could be written compactly in the critical
exponent form as  
\begin{eqnarray} 
\lambda_{+,1}^{C_2(G)}(a_c) &=& 
\left[ [32\mu^5n^2 + 32\mu^5n + 32\mu^5 + 8\mu^4n^4 + 16\mu^4n^3 
- 120\mu^4n^2 - 128\mu^4n \right. \nonumber \\
&& \left. ~~ - 160\mu^4 - 32\mu^3n^4 - 64\mu^3n^3 + 160\mu^3n^2 
+ 192\mu^3n + 316\mu^3 \right. \nonumber \\
&& \left. ~~ + 48\mu^2n^4 + 96\mu^2n^3 - 78\mu^2n^2 - 126\mu^2n - 306\mu^2 
- 31\mu n^4 - 62\mu n^3 \right. \nonumber \\ 
&& \left. ~~ + 31\mu n + 146\mu + 7n^4 + 14n^3 + 7n^2 - 28]  
\Gamma(n+2-\mu)\Gamma(\mu) \right. \nonumber \\ 
&& \left. ~~~ /[8n(\mu - 1)^3(n + 2)(n^2 - 1) 
\Gamma(2-\mu)\Gamma(\mu + n)] \right. \nonumber \\ && \nonumber \\  
&& \left. ~-~ [32\mu^5n^2 + 32\mu^5n + 32\mu^5 - 144\mu^4n^2 - 144\mu^4n 
- 160\mu^4 - 4\mu^3n^4 \right. \nonumber \\ 
&& \left. ~~~~~~ - 8\mu^3n^3 + 240\mu^3n^2 + 244\mu^3n + 316\mu^3 + 16\mu^2n^4 
+ 32\mu^2n^3 \right. \nonumber \\ 
&& \left. ~~~~~~ - 180\mu^2n^2 - 196\mu^2n - 306\mu^2 - 20\mu n^4 - 40\mu n^3 
+ 59\mu n^2 \right. \nonumber \\
&& \left. ~~~~~~ + 79\mu n + 146\mu + 8n^4 + 16n^3 - 6n^2 - 14n - 28] 
\right. \nonumber \\ 
&& \left. ~~~~~~~ /[8n(\mu - 1)^3(n + 2)(n^2 - 1)] 
\right. \nonumber \\ && \nonumber \\  
&& \left. ~+~ 2(\mu - 1)S_1(n) \frac{}{} \right] 
\frac{\mu C_2(G) \eta_1^{\mbox{o}} }{(2\mu-1)(\mu-2)T(R)}  
\nonumber \\  
\label{exponent} 
\end{eqnarray} 
where $S_l(n)$ $=$ $\sum_{r=1}^\infty 1/r^l$, $d$ $=$ $2\mu$, $f^{acd}f^{bcd}$ 
$=$ $C_2(G)\delta^{ab}$ and 
\begin{equation} 
\eta^{\mbox{o}}_1 ~=~ 
\frac{(2\mu-1)(\mu-2)\Gamma(2\mu)}{4\Gamma^2(\mu)\Gamma(\mu+1)\Gamma(2-\mu)} ~. 
\end{equation}  
We recall that a feature of the large $\Nf$ approach to computing information 
on the perturbative coefficients of (\ref{coeffdef}) was that the anomalous 
dimensions of the {\em eigen}-operators of (\ref{mixmat}) at criticality were 
determined and denoted by $\lambda_{\pm}(a_c)$ $=$ $\sum_{i=1}^\infty 
\lambda_{\pm,i}(a_c)/\Nf^i$, \cite{10}. That eigen-operator which was 
predominantly gluonic in content corresponds to the eigen-critical exponent 
$\lambda_+(a_c)$ whilst $\lambda_-(a_c)$ corresponds to the dimension of the 
mainly fermionic eigen-operator. The location of the fixed point, $a_c$, is 
given by the non-trivial zero of the $d$-dimensional QCD $\beta$-function and 
in the present notation, \cite{16,17,18,19,20}, 
\begin{eqnarray}
a_c &=& \frac{3\epsilon}{4T(R) \Nf} ~+~ \frac{1}{T^2(R) \Nf^2} \left[ 
\frac{33}{16} C_A\epsilon ~-~ \left( \frac{27}{16}C_F+\frac{45}{16}C_A \right) 
\epsilon^2 ~+~ \left( \frac{99}{64}C_F + \frac{237}{128}C_A \right) \epsilon^3 
\right. \nonumber \\ 
&& \left. +\, \left( \frac{77}{64}C_F + \frac{53}{128}C_A \! \right) \!  
\epsilon^4 ~-~ \frac{3}{1024} \left[ (288\zeta(3) + 214)C_F 
+ (480\zeta(3)-229)C_A \right] \epsilon^5 \,+\, O(\epsilon^6) \right] 
\nonumber \\ 
\label{critcoup} 
\end{eqnarray}
to $O(1/\Nf^3)$ in the large $\Nf$ expansion where $d$ $=$ $4$ $-$ $2\epsilon$
and $\zeta(r)$ is the Riemann zeta function. From (\ref{exponent}) and 
(\ref{critcoup}), we find, for instance, 
\begin{eqnarray} 
d_{51}^{C_2(G)} &=& \frac{128[9n^4 + 18n^3 + 79n^2 + 70n + 32]S_3(n)} 
{243(n + 2)(n + 1)^2(n - 1)n^2} \nonumber \\  
&& +~ \frac{256[9n^4 + 18n^3 + 79n^2 + 70n + 32]S_1^3(n)}  
{243(n + 2)(n + 1)^2(n - 1)n^2} \nonumber \\  
&& -~ \frac{128[63n^6 + 189n^5 + 821n^4 + 1327n^3 + 1176n^2 + 544n + 96]  
S_1^2(n)}{243(n + 2)(n + 1)^3(n - 1)n^3} \nonumber \\  
&& -~ 256[4n^{10} + 20n^9 + 17n^8 - 52n^7 - 414n^6 - 976n^5 - 1521n^4 - 1516n^3
\nonumber \\ 
&& ~~~~~~~~ - 930n^2 - 320n - 48]S_1(n)/[243(n + 2)(n + 1)^4(n - 1)n^4] 
\nonumber \\ 
&& -~ \frac{2048[n^2 + n + 1]\zeta(4)}{27(n + 2)(n^2 - 1)n} ~+~ 
\frac{1024S_1(n)\zeta(4)}{27} ~-~ \frac{10240S_1(n)\zeta(3)}{243} \nonumber \\  
\nonumber \\  
&& +~ \frac{128[3n^6 + 9n^5 + 307n^4 + 599n^3 + 746n^2 + 448n + 96]\zeta(3)}  
{243(n + 2)(n + 1)^2(n - 1)n^2} \nonumber \\  
&& +~ 4[155n^{12} + 930n^{11} + 1455n^{10} - 1250n^9 - 9879n^8 - 21786n^7 
\nonumber \\ 
&& ~~~~~~ - 47107n^6 - 80550n^5 - 97392n^4 - 78336n^3 - 40000n^2 \nonumber \\
&& ~~~~~~ - 11776n - 1536]/[243(n + 2)(n + 1)^5(n - 1)n^5] ~.  
\end{eqnarray} 
This illustrates earlier remarks in that this coefficient clearly has a fifth 
order pole at $n$ $=$ $0$ which is a degree larger than the previous loop 
order, \cite{10}. Clearly to all orders this degree of divergence will 
increase. However, since (\ref{exponent}) encodes the structure of this 
particular coefficient of $\gamma_{gg}(a)$ to all orders in the coupling 
constant in leading order in $1/\Nf$ as a function of $n$, we can examine the 
form of the exponent for $n$ $=$ $0$. In other words prior to making the 
connection with explicit perturbation theory. Therefore setting $n$ $=$ $0$ in 
(\ref{exponent}) we can determine if the expression is divergent which would 
indicate that such a pole persists in $\gamma_{gg}(a)$ or if it is finite. The 
latter case would indicate consistency with the {\em supposition} from more 
general principles, \cite{4,8}. First, by inspection of (\ref{exponent}) the 
$S_1(n)$ independent terms each have simple poles in $n$. However, by careful 
examination of the numerators of each term and in particular their  
$n$-independent terms, it is evident that the residue at $n$ $=$ $0$ is the 
same for both whilst the relative sign between the terms means that overall the
residue is zero for the simple pole at $n$ $=$ $0$. Also for the term involving
$S_1(n)$ one writes 
\begin{equation} 
S_{l+1}(n) ~=~ \frac{(-1)^l}{l!} \left[ \psi^{(l)}(n+1) ~-~ \psi^{(l)}(1) 
\right]  
\end{equation} 
for general $l$, which vanishes in the limit $n$ $\rightarrow$ $0$ where 
$\psi(x)$ is the derivative of the logarithm of the Euler $\Gamma$-function. 
Hence the exponent (\ref{exponent}) is in fact finite at $n$ $=$ $0$. This is 
consistent with the resummed splitting function being non-singular at this 
point. However, it is important to recall that we have only demonstrated this 
property for that part of the series which is {\em leading} order in the 
$1/\Nf$ expansion. Moreover that this part is not inconsistent with general 
considerations is, we believe, an important observation and the central result 
of this article. Furthermore, we record that when $n$ $=$ $0$ the exponent 
(\ref{exponent}) is 
\begin{eqnarray} 
\left. \lambda_{+,1}^{C_2(G)}(a_c) \right|_{n=0} &=& 
\left[ \frac{}{} [16\mu^4 - 64\mu^3 + 94\mu^2 - 59\mu + 14](\mu-1)(\mu-2) 
[ \psi(\mu-1) - \psi(3-\mu)] \right. \nonumber \\ 
&& \left. ~+~ \mu(8\mu^4 - 26\mu^3 + 23\mu^2 - 4) \frac{}{} \right]  
\frac{\mu C_2(G) \eta_1^{\mbox{o}}}{8(2\mu-1)(\mu-1)^3(\mu-2)^2} ~. 
\end{eqnarray}  

For completeness, we also comment on the $n$ $\rightarrow$ $0$ limit of the 
critical exponents of the other twist-$2$ operators at $O(1/\Nf)$. We recall 
that, \cite{12},  
\begin{eqnarray}
\lambda_{-,1}(a_c) &=& \frac{4\mu(\mu-1)C_2(R)\eta^{\mbox{o}}_1}{(2\mu-1) 
(\mu-2)T(R)\Nf} \left[ \frac{(\mu-1)(n-1)(2\mu+n-2)}{2\mu(\mu+n-1)(\mu+n-2)}
{}~+~ [\psi(\mu-1+n) - \psi(\mu)] \right. \nonumber \\
&&-~ \left. \frac{\Gamma(n-1)\Gamma(2\mu)}{4(\mu+n-1)(\mu+n-2)
\Gamma(2\mu-1+n)} \right. \nonumber \\ 
&&~~~~ \times \left. \left[ (n^2+n+2\mu-2)^2 + 2(\mu-2)(n(n-1)(2\mu-3+2n) 
+ 2(\mu-1+n)) \right] \frac{}{} \right] \nonumber \\ 
\label{fermdim} 
\end{eqnarray}
for the flavour singlet fermionic operator where $T^a T^a$ $=$ $C_2(R)$. The 
critical exponent for the non-singlet operator dimension corresponds to the 
first two terms of (\ref{fermdim}) and it is clear that in the limit we are 
interested in that that exponent is finite. In particular 
\begin{equation}
\left. \lambda_{-,1}^{\mbox{\footnotesize{non-singlet}}}(a_c) \right|_{n=0}
{}~=~ -~ \frac{4(2\mu^2-4\mu+1)C_2(R)\eta^{\mbox{o}}_1} 
{(2\mu-1)(\mu-2)^2T(R)\Nf} ~.  
\end{equation}
By contrast, the singlet fermionic eigen-operator, which in perturbation theory
corresponds to the combination of coefficients $(a_{l1} - b_{l1}c_1/d_{11})$ at
$l$th loop, diverges at $n$ $=$ $0$ having the behaviour
\begin{equation}  
\lambda_{-,1}(a_c) ~ \sim ~ \frac{4\mu(\mu-1)(2\mu-3)C_2(R)\eta^{\mbox{o}}_1}
{n(\mu-2)^2 T(R) \Nf} 
\end{equation}  
as $n$ $\rightarrow$ $0$ with the singularity arising from the final term of 
(\ref{fermdim}) which corresponds to Feynman diagrams with the operator 
inserted in a closed quark loop.

To conclude with we have demonstrated that for unpolarized scattering the 
behaviour of $\gamma_{gg}(a)$ at $n$ $=$ $0$ at leading order in the large 
$\Nf$ expansion is consistent with the function being finite at this point. 
However, it is worth putting this result in context with other observations 
from the BFKL formalism which has also been used to study the anomalous 
dimensions discussed here. In \cite{21} it was shown that the $C_2(G)$ sector 
of $\gamma_{gg}(a)$ in the BFKL approach not only does not contain any poles in
$n$ but has in fact three branch points. Indeed this does not appear to be just
a feature of QCD. If one examines the toy model of scalar $\phi^3$ theory in 
six dimensions the {\em complete} analogous anomalous dimension was resummed at
leading order without reference to the small $x$ limit and again only branch 
points and no poles were observed, \cite{22}. For other anomalous dimensions in
QCD, the non-singlet and polarized dimensions also do not have poles at $n$ $=$
$0$ but branch points, \cite{23}. (For a recent review of these issues see, for 
example, \cite{24}.) Therefore it is reassuring that our large $\Nf$ 
observation is also not inconsistent with the BFKL formalism. In light of this 
it would be interesting to go beyond our leading order $1/\Nf$ analysis to 
confirm the absence of the $n$ $=$ $0$ pole at next order as well as being able
to understand the branch point structure in the large $\Nf$ point of view. To 
compute the $O(1/\Nf^2)$ correction to $\lambda_+(a_c)$, though, would involve 
the evaluation of a large set of $1/\Nf$ graphs which contain, for example, 
several six loop diagrams and is therefore, we believe, not attainable in the 
near future. 

\vspace{1cm} 
{\bf Acknowledgement.} This work was carried out with the support of 
{\sc PPARC} through a Postgraduate Studentship (JFB) and an Advanced Fellowship
(JAG).

\end{document}